\documentclass[a4]{aa}
 
\usepackage{graphicx}
 
\newcommand{\C}{\v{C}erenkov }
\newcommand{\W}{\hbox{$\bar{w}$} }
\newcommand{\dis}{{\em dis }}

\begin{document}
 
\title{A new {\it template} background estimate for source searching in TeV $\gamma$-ray astronomy}
 
\titlerunning{A Template Background for TeV Astronomy}

\author{G.P. Rowell}

\institute{Max Planck Institut f\"ur Kernphysik,
         Postfach 103980, D-69029,
         Heidelberg, Germany}

\authorrunning{Rowell}  

\date{Received 21 May 2003 / Accepted 30 July 2003}
 
\offprints{G.P. Rowell\\
\email{Gavin.Rowell@mpi-hd.mpg.de}}

\abstract{
A new method is described that permits quickly and easily, a 2-dimensional search for TeV $\gamma$-ray
sources over large fields of view ($\sim 6^\circ$) with instruments utilising the imaging atmospheric \C 
technique. It employs as a background estimate, 
events normally rejected according to a cosmic-ray background rejection criterion based on image shape, 
but with reconstructed directions overlapping the source of interest. 
This so-called {\em template} background model is demonstrated using example data taken with the 
stereoscopic HEGRA System of \C Telescopes. Discussion includes comparisons with a conventional
background estimate and limitations of the model. The template model is well suited 
to the search for point-like, moderately extended sources and combinations thereof, and compensates well
for localised systematic changes in cosmic-ray background response.
}

\maketitle

\section{Motivation \& Introduction}
The search for new astrophysical sources of TeV $\gamma$-ray emission today is well
motivated by the multi-wavelength picture of candidate sources, 
and also by the high performance offered by present and future instruments.
Many of the $>$150 unidentified EGRET sources (MeV - GeV energies) have positional uncertainties
approaching several degrees (Hartman et al. \cite{Hartman:1}), presenting exciting opportunities for
TeV instruments to identify possible counterparts over the coming years.
The ground-based imaging atmospheric \C technique presently offers on-axis flux
sensitivities better than (for energies $>1$ TeV) $\sim 10^{-12}$ erg cm$^{-2}$ s$^{-1}$ for 50
hrs exposure, and can achieve arc-minute scale angular resolution over fields of view 
approaching 5$^\circ$ diameter (see e.g. Konopelko et al. \cite{Konopelko:2}). Rather wide surveys 
for new TeV sources are therefore possible even
from singly pointed observations. Techniques to generate 2D {\em skymaps} of event excess significance 
for such surveys have therefore become more important in recent years. Critically important, for such skymap 
generation is an estimate of the cosmic ray background over the field of view (FoV).
Many expected, and known sites of TeV $\gamma$-ray
production, such as supernova remnants, pulsars/plerions, microquasars, 
and nearby extragalactic sources will likely present complicated morphology for future 
instruments as sensitivities improve. Particularly in cases where
combinations of such sources might be expected in the FoV, it is vital that the background estimate is not
contaminated by other TeV sources and various systematic biases are well understood.
 
Described here is the {\em template} background model (an earlier version 
is described in Rowell \cite{Rowell:1}), designed to provide a background estimate 
for sources of interest at all positions in the
FoV. The model is demonstrated on archival data from the HEGRA
System of \C telescopes (HEGRA CT-System). For detailed descriptions of this instrument and its performance
see Daum et al. \cite{Daum:1} and P\"uhlhofer et al. \cite{Puehl:1}. 
Comparisons are made between results of the template model and those taken 
from a conventional background model presently in use in HEGRA CT-System data analysis. Some features and
limitations of the template model are described.

\section{Conventional Background Estimates and 2D Skymap Generation}

\begin{figure}[t]
 \begin{center}
 \includegraphics[width=9.0cm]{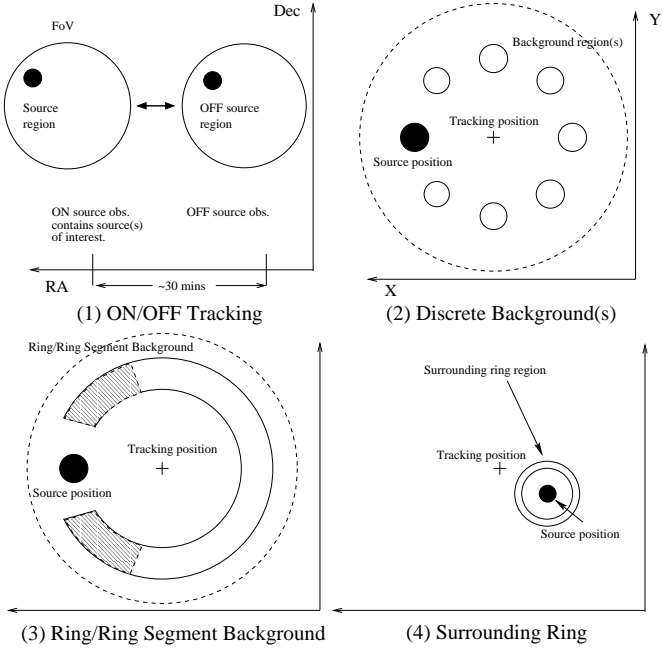} 
 \end{center}
 \caption{Illustration of some {\em conventional} background estimation methods. (1)
          Separate OFF source observations are taken at matching zenith angles
          to those of ON source data. (2) Discrete background regions within the same FoV at a
          common source-to-tracking distance so as to match background response. (3) A continuous version of
          (2), ring or ring segments
          may be used instead. Regions adjacent to the source (shaded) may be preferred to
          avoid global systematic changes in response. (4) A ring region completely surrounding the source.
          Methods 2,3 \& 4 are termed displacement background models. Method 4 is used when
          the source is too close to the tracking position to implement methods 2 \& 3, or over
          the entire FoV.}
 \label{fig:backgrounds}
\end{figure}

Present ground-based instruments utilising the imaging atmospheric \C technique (100 GeV to $\sim$30 TeV), 
must detect $\gamma$-ray initiated events against a background of vastly outnumbering isotropic 
cosmic-ray (CR) initiated events. First, a set of so-called $\gamma$-ray cuts are employed to
preferentially select events conforming to a $\gamma$-ray
hypothesis. These are {\em a-priori} chosen, and based primarily on \C image shape parameters, for 
example {\em width}, {\em length} (Hillas \cite{Hillas:1}), and {\em mean-scaled-width} (\W) 
(Konopelko \cite{Konopelko:1},  Daum et al. \cite{Daum:1}).  
Second, event arrival directions are also reconstructed based on
the orientation of recorded \C images (see eg. Buckley et al. \cite{Buckley:1}, 
Hofmann et al. \cite{Hofmann:1}) and a so-called directional cut made on the 
distance $\theta$ (or analogous parameters) between the reconstructed and assumed arrival directions.  
To assess the statistical 
significance of any region in the FoV, an estimate of the CR background surviving both $\gamma$-ray and
directional cuts must be made.
This background estimate can be derived from separate observations of different OFF tracking positions
(ON/OFF tracking), or control regions displaced within the same FoV containing the source region.
The latter types of background models are termed here displacement backgrounds, and a number of different  
displacement background geometries are available.
Background regions are chosen so that their CR response matches as closely 
as possible, that of the source region (see Fig.~\ref{fig:backgrounds}). 
In ON/OFF tracking this is 
achieved by choosing background regions with the same relative displacement from respective 
tracking positions as that for the source. For displacement backgrounds, such regions are 
mirrored through the tracking position (methods 2 \& 3), except for sources very close to the 
tracking position in which a region surrounding completely the source can be implemented (method 4).
In this latter case an extra correction is applied to account for differences in CR response. 
Provided the FoV is large compared to the angular resolution and source size, and global systematic
changes in CR response are minimal or can be alleviated, displacement backgrounds may be implemented. Displacement
backgrounds are favoured over the ON/OFF method since they do not require extra time taking dedicated
OFF source observations. A recent modification of the ON/OFF method, however, is proposed by Petry \cite{Petry:1}.
The large cameras ($\sim$4.3$^\circ$ diameter) and stereoscopic trigger employed by the HEGRA CT-System 
achieves a quite flat FoV (FWHM $\sim3^\circ$) that is amenable to displacement background models 
over the entire FoV. 
In this paper, using example data from the HEGRA CT-System, results from a combination of displacement methods 
3 and 4 (termed here the {\em displacement} background model) are used to compare with the template model. 
For source to tracking distances larger than $\sim 0.4^\circ$, a ring segment or arc (method 3 ) 
adjacent to the source region (within position angles 135$^\circ$ to 165$^\circ$) is used, 
otherwise, method 4 is chosen. 
Ring background regions rather adjacent to the source are often preferred to avoid global 
systematic changes (of $\sim$few percent) in CR response across the FoV. 

For a given region in the FoV, the source events $s$ are derived from within a nominal source region 
according to some directional cut $\theta<\theta_{\rm cut}$ and $\gamma$-ray cuts on image shape parameters. 
The number of background events $b$ are
derived according to the above-described displacement background methods, after also applying the same 
$\gamma$-ray cuts.
The statistical significance $S$ of excess counts $s-\alpha b$ is then estimated from
 Eq. (17) of Li \& Ma (\cite{Li:1}):
\begin{equation}
S = \sqrt{2} \left( s\,{\rm ln}\left[\frac{s(1+\alpha)}{\alpha(s+b)}\right] \, 
                          + b\,{\rm ln}\left[\frac{b(1+\alpha)}{s+b}\right] \right)^{\frac{1}{2}}
 \label{eq:signif}
\end{equation}
where a normalisation $\alpha$ accounts for the solid angle ratio between the source
and background regions when displaced
backgrounds are used, or different observation times if ON/OFF tracking
is used. Under the displaced background models applied to HEGRA CT-System
data, typical normalisation values of 0.05 to 0.3 are achieved,
thereby considerably improving the background estimate compared to an $\alpha=1.0$ situation
usually achieved in ON/OFF tracking.

In 2D skymap generation, the above procedure is repeated over
the FoV at a series of grid positions. Directional parameters such
as $\theta$ and those necessary to form the background estimate are then re-calculated with respect to each 
grid position, and excesses and significances are calculated accordingly. 

\section{The Template Background Model}
\label{sec:model}

The template background model invokes a different philosophy to that
of displacement background models. Instead of selecting spatially different
regions in the FoV for background estimates, template model background
events are comprised of those with their
arrival directions reconstructed in the same region as that for the
source, yet separated in image shape parameter space. Such events can 
form a suitable {\em template} of response for $\gamma$-ray-like CR events over the FoV. 

\begin{figure}[t]
 \begin{center}
 \includegraphics[width=7.0cm]{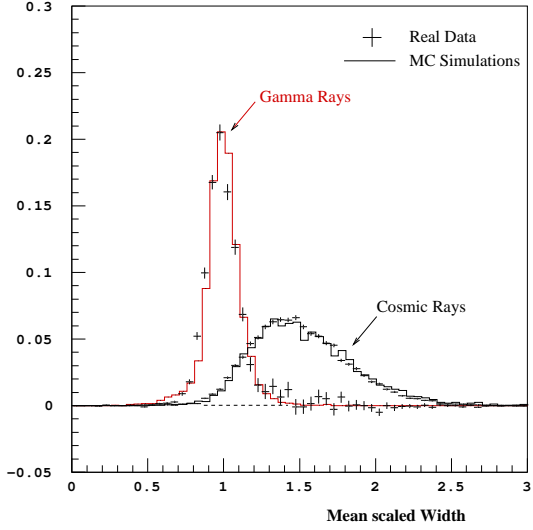}
 \end{center}
 \caption{Normalised distributions of \W for $\gamma$-ray and cosmic-ray (CR) events, based on Monte
 Carlo simulations (solid histograms) and real data (error bars). Adapted from Konopelko \cite{Konopelko:2}.}
 \label{fig:MSW_dist}
\end{figure}

As applied to data of the HEGRA CT-System, the shape parameter {\em mean-scaled-width} (\W) is used
to construct the template background estimate. \W is the image {\em scaled-width}, averaged over all
images accepted for analysis. The scaled-width is the image {\em width} scaled according to a $\gamma$-ray 
hypothesis. This scaling is dependent upon the total photoelectron
yield of the image, the zenith angle of observations, and the impact parameter or distance to the
air-shower core. By design (see Fig.~\ref{fig:MSW_dist}), true $\gamma$-ray events 
have \W values centred at unity, leaving the generally broader CR events to fill higher
values. 
In the search for weak, background-dominated sources in CT-System data, the cut \W$<$1.1 has been found optimal. 
Template 
background events should be derived from a \W regime containing little or no true $\gamma$-rays, i.e. \W$>$1.1.
Experience has found that events within the regime $1.3<$\W$<1.5$ work well, although regimes containing
higher \W values have proved adequate.
To apply the template model practically however, two dominant systematic effects must be corrected:
\begin{itemize}
 \item {\bf Radial Correction}: Differences in the radial CR acceptance between the two \W regimes.
 \item {\bf Zenith-Correlated Correction}: Differences in CR acceptance between the two \W regimes
        which correlate with the zenith angle of reconstructed event directions.
\end{itemize} 
For a nominal region in the FoV defined by a cut $\theta<\theta_{\rm cut}$, $\gamma$-ray-like 
events $s$ are selected according to \W$<$1.1.
Similarly after applying the {\em same} cut on $\theta$, template background events $b$ are defined as those 
satisfying the condition $1.3<$\W$<1.5$. In summary template background events are assigned a weight $\omega_i$:
\begin{eqnarray}
 s = \sum_{i} \omega_i & \omega_i = 1.0 & \,\, (\bar{w}<1.1)\\
 b = \sum_{i} \omega_i & \,\, \omega_i = \frac{\textstyle p_s(\theta^2_{\rm track,i})}{\textstyle p_b(\theta^2_{\rm track,i})} \, f(\Delta {\rm z_i}) & 
                        \,\,   (1.3<\bar{w}<1.5) \label{eq:bcorrz}
\end{eqnarray}
for events $i$ in the source region. The correction terms applied to $b$ can be
described with suitable parametrisations as follows. The radial acceptance of the $\gamma$-ray-like (\W$<$1.1) 
and template ($1.3<$\W$<1.5$) regimes over the entire FoV are sufficiently fitted out to radius $R$,
with 8th-order polynomials, $p_s$ and
$p_b$ respectively and normalised. Both polynomials are functions of
the event distance to tracking position $\theta_{\rm track}$. A strong increase or pile-up of reconstructed
event directions due to camera edge effects is normally seen at the FoV edge. 
Ignoring \C images whose centroid distance from the camera centre \dis exceeds 1.7$^\circ$ effectively
suppresses this pile-up, and allows fitting of the FoV response to the physical camera limits.
The choice of $R$ is limited by event statistics, and practically, 
can reach values up to $3^\circ$. Fitting such polynomials assumes that the
radial response is azimuthally symmetric. However, a second systematic effect,  
a gradient $f(\Delta z)$ correlating with the difference in zenith angle $\Delta z$ between the event 
and tracking position, is also generally present. This can be described by a linear
function vs. $\Delta z$, of the ratio of FoV responses between the $\gamma$-ray-like and template
\W regimes. This linear fit is made after first correcting for the radial profiles. Thus the 
template correction functions are generated in a two-step process (Appendix~\ref{sec:appendixa} summarises
these corrections for a particular example discussed shortly). 
The radius $R$ also defines the FoV used in determining the overall normalisation ratio $\alpha$ when 
estimating event excess and significance (e.g. according to Eq. (\ref{eq:signif})). If the total number of 
events in each \W regime out to a radius $R$ is given by $P_s$ and $P_b$ respectively, the template 
normalisation $\alpha$ is given by: 
\begin{equation}
 \alpha = P_s / P_b
  \label{eq:norm}
\end{equation}  
\begin{figure}[ht]
 \begin{center}
 \includegraphics[width=8.0cm]{fov2.epsi}
 \end{center}
 \caption{Illustration of the template background geometry with the
  FoV and normalisation region (defined by radius $R$) in
  relation to various source positions. The template background and `$\gamma$-ray-like' events are defined
  by the \W (MSW) regimes 1.3$<$\W$<1.5$ and \W$<$1.1 respectively. Shown here un-normalised, the template
  regime contains a factor 3$\sim$4 more events than that of the $\gamma$-ray-like regime.
  Source regions of interest are defined by 
  a cut on some directional parameter, e.g. $\theta<\theta_{\rm cut}$, and are spatially coincident 
  in both \W regimes.  
  The normalisation region should be sufficiently large to encompass all such source regions.}
 \label{fig:templateidea}
\end{figure}
Alternatively, one may also integrate out to $R$, the polynomial functions (un-normalised) 
$p_s$ and $p_b$, to obtain a similar result. Typical $\alpha$ values achieved 
are quite similar to that of the displacement background model. The template model philosophy is illustrated in
Fig.~\ref{fig:templateidea}. 
A caveat is that the normalisation region must 
encompass all source regions. This will set
a natural limit on the size of any source region, although systematic effects (section~\ref{sec:limitations})
usually dominate for large source sizes. Note that the template model correction terms generated from a 
given dataset should strictly be applied only to these same data.

\section{Template Model Performance on Real Data}
 \label{sec:realdata}

2D skymaps of excess significance are shown in Fig.~\ref{fig:skymaps}, illustrating the template model 
performance
at various stages of correction. About 50 hours of CT-System exposure on Tycho's SNR out 
to a radius $R=2.24^\circ$ ($\sqrt{5.0}^\circ$) were used in generating these skymaps. Previous analysis of these
data are described in Aharonian et.al \cite{Aharonian:2}.
\begin{figure*}
 \begin{center}
 \includegraphics[width=14.0cm]{tychomap_cr.epsi}

 \includegraphics[width=14.0cm]{tychomap_crc.epsi}

 \includegraphics[width=14.0cm]{tychomap_crcz.epsi}
 \end{center}
 \caption{2D Skymaps  and 1D distributions of significance $S$
          using the template background estimate after various correction levels. 
          (a) no correction,  $w_i=1.0$ 
          (b) radial correction only, $f(\Delta {\rm z_i})=1.0$
          (c) full correction as per Eq. (~\ref{eq:bcorrz}). White star: Nominal
    position (SIMBAD) of Tycho's SNR, White dots: tracking positions, 
    Dashed line: Gaussian ($\mu$=0.0, $\sigma$=1.0), Solid line: Fitted Gaussian.
    Events are summed within a circle of radius 0.12$^\circ$
    at each grid position (120x120 grid), in 0.05$^\circ$x0.05$^\circ$ steps.}
 \label{fig:skymaps}
\end{figure*}
At each grid point in $0.05^\circ$ steps, a $\theta<0.12^\circ$
cut defines both the source (which is applicable for point-like sources) and template background regions. The 
significance $S$ is estimated using Eq. (~\ref{eq:signif}). It is apparent the Tycho skymap statistics
approach that of a Gaussian distribution
only after correcting for both the radial and zenith-correlated systematics. 
A quite strong zenith-correlated gradient of 9\% per degree is present, which if not corrected, leads to
biases in the excess significance peaking at the declination edges of the FoV 
(Appendix~\ref{sec:appendixa} summarises these template correction 
terms in more detail.). The bin statistics in these skymaps are somewhat correlated, and lead
to a slight overestimate in the width $\sigma$ of the 1D distribution of significances. The level of this 
overestimate is taken from a 2D skymap of significances from bins that have no overlap with 
neighbouring bins. The 1D distributions of significances for this skymap is $\sigma=0.93\pm0.03$, implying 
that the template background is slightly over-corrected, effectively underestimating slightly the 
significance in the FoV. Overall one may conclude however that the template model is quite suitable for point source 
searches over the FoV in these data. 
In this example, the minimum
number of telescopes required for stereoscopic direction reconstruction and \W calculation has been chosen at
$n_{\rm tel}\ge 2$, and the stereo reconstruction algorithm \#1 of Hofmann et al. (\cite{Hofmann:1}) was used.
The template model performs equally well however under higher choices of $n_{\rm tel}$ (up to 5),
varying choices of FoV/normalisation radius $R$, and various reconstruction algorithms.

The template model was also tested with a number of {\em a priori} 
chosen positions in the FoV corresponding to established $\gamma$-ray sources.
Presented in Table~\ref{tab:results} are the excess significances for a representative
sample of TeV $\gamma$-ray sources with a variety of  
signal to CR background ratios. Results using background estimates from both the template and 
displacement background models are given for comparison.
In all cases a normalisation radius for the
template background model of $R=2.24^\circ$ and minimum number of images $n_{\rm tel}\ge 2$ have been used.
\begin{table*}
 \begin{center}
{
%
%
 \begin{tabular}{lcccccccr} \hline \hline
  Source  & RA [hr]$^a$ & Dec [deg]$^a$ & $\theta_{\rm cut}$ [deg] &  $s$ & $b$ & $\alpha$ & $s-\alpha b$ & $S$ [$\sigma$]$^b$ \\ \hline \\
  \multicolumn{9}{c}{----- Template background (full correction: Eq (~\ref{eq:bcorrz})) -----}\\
  Cas-A       & 23.390 & 58.82  & 0.12 & 1364 & 4649 & 0.258 & {\bf 165}   & {\bf +4.2} \\ 
  H1426+428   & 14.476 & 42.67  & 0.12 & 385  &  954 & 0.295 & {\bf 104}   & {\bf +5.1} \\  
  1ES1959+650 & 20.000 & 65.15  & 0.12 & 571  & 2236 & 0.210 & {\bf 101}   & {\bf +4.1} \\
  Crab        &  5.576 & 22.01  & 0.12 & 1756 &  884 & 0.282 & {\bf 1507}  & {\bf +48.9} \\ \\
  \multicolumn{9}{c}{----- Displacement background -----}\\         
  Cas-A       & &        & 0.12 & 1364 & 6092     & 0.197 & {\bf 164}   & {\bf +4.2} \\
  H1426+428   & &        & 0.12 & 385  &  997     & 0.283 & {\bf 103}   & {\bf +5.1} \\ 
  1ES1959+650 & &        & 0.12 & 571  & 2437     & 0.201 & {\bf 81}    & {\bf +3.2} \\ 
  Crab        & &        & 0.12 & 1756 & 1281     & 0.226 & {\bf 1466}  & {\bf +48.2}  \\ \hline \hline
  \multicolumn{9}{l}{\scriptsize a. J2000 epoch.}\\
  \multicolumn{9}{l}{\scriptsize b. Statistical significance from Eq. (~\ref{eq:signif}).}\\
 \end{tabular}
}
 \end{center}
 \caption{Event excesses $s-\alpha b$ and significances $S$ derived using the template and displacement
          background models for a number of established TeV $\gamma$-ray sources.
          The source region is defined by the cut $\theta<\theta_{\rm cut}$.
          $s$: $\gamma$-ray-like events; $b$: background estimate; $\alpha$: normalisation.
          Datasets are from the HEGRA CT-System archive; Cas~A (212 hours), 
          H1426+428 (42 hours, 1999/2000), 1ES1959+650 (94 hours, 2000/2001),
          Crab (33 hours).}
 \label{tab:results}
\end{table*}
The template model provides derived excesses $s-\alpha b$ quite consistent with those obtained from the
displacement model. Since the example sources of
Table~\ref{tab:results} lie quite close to their respective tracking positions,
the zenith-correlated correction term of the template background has little effect, and hence results
only for the fully corrected template background are shown.
For the Crab, a slight modification regarding the template correction is made to the fit limits of the
polynomials $p_s$ and $p_b$, to exclude regions containing 
this strong source. The polynomial fits are affected by strong sources near ($<0.5^\circ$) the tracking position
and these should be avoided.
It should be stressed that the derived excesses and significances for
the Cas~A and H1426+428 and 1ES1959+650 
here are generally lower then those of presently published 
analyses (Aharonian et al. \cite{Aharonian:1,Aharonian:3}, Horns et al. \cite{Horns:2}), 
which employed the more sensitive reconstruction algorithms \#2 and \#3
(Hofmann et al. \cite{Hofmann:1}), and in some cases a higher minimum $n_{\rm tel}$.
The template model does reproduce published results when employing the corresponding analyses. 

Finally, a particularly useful feature of the template model is demonstrated against a  
problematic systematic effect arising when camera pixels view bright stars in the FoV.
Increased noise fluctuations (and corresponding anode currents) in camera pixels viewing bright stars 
can distort derived image parameters.
Affected pixels, usually the same for all five cameras, are therefore removed
from a camera trigger dynamically during data taking and also from image analysis at a software level. 
Any pixel with anode current $\geq3 \mu$A is not considered in estimation of image parameters and consequent
event direction reconstruction (Bulian et al. \cite{Bulian:1}, Daum et al. \cite{Daum:1}). 
A residual effect of this latter step is to leave  
a systematic deficit of reconstructed event directions in a region broadly centred on the offending star.
This effect was verified by artificially removing certain pixels before image parametrisation.
In practise, up to two to three pixels may be removed during data taking at any given time depending on the 
star brightness. The Crab field, with $\zeta$-Taurii ($m_{\rm B}$=2.9)
and the Cas-A field with AR-Cas ($m_{\rm B}$=4.8),
are ideal examples to illustrate such star {\em deficits}.
Quantitatively, star deficits amount to a
reduction in CR events up to $\sim$20\% over a diameter $\sim0.2^\circ$. Table~\ref{tab:startest} summarises 
the template (after full correction only) and displacement background model performance at the  
$\zeta$-Taurii and
AR-Cas positions. Since the template background closely follows the CR background response of gamma-ray-like
events, star deficits are considerably reduced when
using the template model. The displacement background however, which does not include the
star region, produces a strong negative bias. Conversely, if displacement
background regions (applied to another source region) were to overlap star deficit(s), some positive
biases could be introduced. A large number of bright stars in the FoV would in general
render application of displacement background models problematic. 
\begin{table*}
 \begin{center}
{
 \begin{tabular}{lccccccccr} \hline \hline
  Star & RA [hrs]$^a$ &  Dec [deg]$^a$  &  $m_{\rm B}^b$& $\theta_{\rm cut}$ [deg] &  $s$ & $b$ & $\alpha$ & $s-\alpha b$ & $S$ [$\sigma$]$^c$  \\ \hline \\
  \multicolumn{10}{c}{----- Template background (full correction: Eq. (~\ref{eq:bcorrz})) -----}\\
  $\zeta$-Taurii & 5.627 & 21.14 & 2.4 &        0.15 &  970 & 3399 & 0.290  & {\bf $-$16}  & {\bf $-$0.4}   \\
  AR-Cas         & 23.501 & 58.55 & 4.8 &       0.15 & 1294 & 5293 & 0.258  & {\bf $-$72} & {\bf $-$1.7}   \\ \\
  \multicolumn{10}{c}{----- Displacement background -----}\\
  $\zeta$-Taurii && & 2.4 &        0.15 &  970 &  6241 & 0.201 & {\bf $-$290} & {\bf $-$7.6}   \\
  AR-Cas         & & & 4.8 &       0.15 & 1294 &  7358 & 0.213 & {\bf $-$273} & {\bf $-$6.4}   \\ \hline \hline
  \multicolumn{10}{l}{\scriptsize a. J2000 epoch.}\\
  \multicolumn{10}{l}{\scriptsize b. B-band star magnitude.}\\
  \multicolumn{10}{l}{\scriptsize c. Statistical significance from Eq. (~\ref{eq:signif}).}\\
 \end{tabular}
}
 \end{center} 
 \caption{Event excesses $s-\alpha b$ and significances $S$ at star positions within 
            the Crab ($\zeta$-Taurii, 161 hrs) and Cas-A (AR-Cas, 212 hrs) fields of view.
          Compared are results using the template and displacement
          background models.} 
 \label{tab:startest}
\end{table*}
Overall the template model appears able to match localised systematic changes in 
response of $\gamma$-ray-like CR events in the FoV.

\section{Limitations}
 \label{sec:limitations}
The template model as designed here employs corrections only for large-scale systematic changes in CR response, 
correlating radially from the tracking position and with event zenith angle. 
Higher-order systematics (correlating with alternate system parameters) arising from, say, 
differences in telescope camera response can also reasonably be expected, despite long-term efforts 
to maintain a homogeneous response in the HEGRA CT-System (see P\"uhlhofer et al. \cite{Puehl:1}). 
Since systematic effects will grow in proportion to the size of the source region and also exposure, 
these are better-revealed in deep exposure datasets. 
For the Tycho's SNR dataset however, significance skymaps generated with larger source sizes in mind
(e.g. $\theta_{\rm cut}$ up to 0.6$^\circ$) do not show behaviour significantly differing from a Gaussian
distribution.
An upper limit to the systematic error in the derived excess $s-\alpha b$ across the Tycho FoV 
can be estimated by using subjectively-chosen `hot' and `cold' regions (of size $\sim1.0^\circ$) 
of the fully corrected skymap. In this way
a systematic error of $\sim \pm4$\% in units of the normalised background $\alpha b$ is calculated. 

\begin{figure*}
 \begin{center}
 \includegraphics[width=15.0cm]{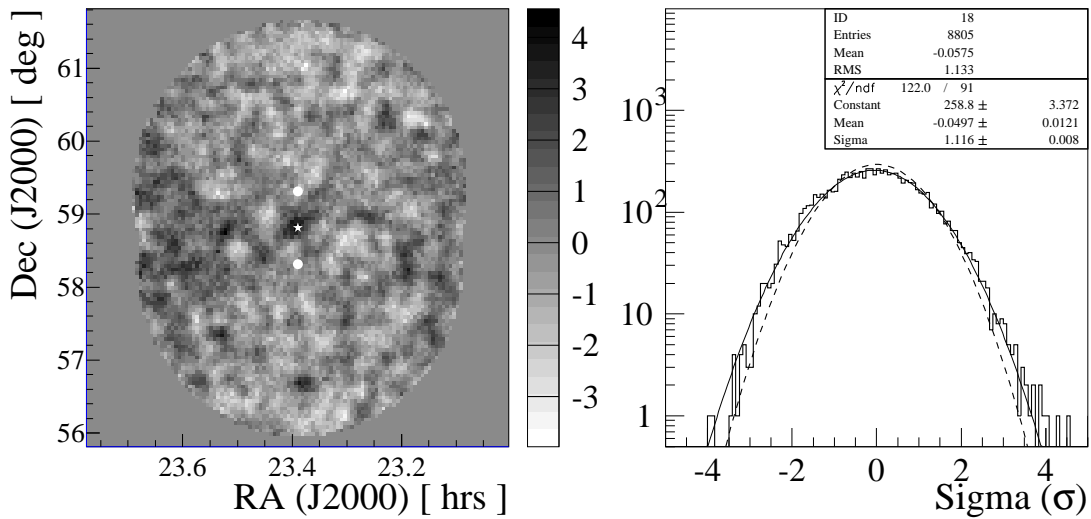}
  \end{center}
 \caption{Skymap (2D \& 1D) of significance $S$ 
          using the fully corrected template background estimate for Cas-A data. 
          White star: Nominal
    position (SIMBAD) of Cas-A, White dots: tracking positions. 
    Dashed line: Gaussian ($\mu$=0.0, $\sigma$=1.0). 
    Details for skymap construction are the same as for Fig~\ref{fig:skymaps}.}
 \label{fig:cas-a_skymap}
\end{figure*}

The Tycho's SNR dataset is representative of many in the CT-System archive of order 50 hrs exposure,
with $\geq$4 telescopes of the CT-System in operation.
Such an exposure is generally found not sufficient to reveal higher-order systematic effects.
The Cas~A dataset comprises a deeper exposure ($>200$ hrs).
The fully corrected template model skymap of these data (Fig~\ref{fig:cas-a_skymap}) shows
a strong deviation from Gaussian behaviour (1D $\sigma=1.126\pm0.008$) already when searching for point
sources ($\theta_{\rm cut} < 0.12^\circ$). A `hotter' group of bins is seen at RA$>23.5$ h, Dec 
58$^\circ$ to $59^\circ$, representing an additional systematic excess of order few percent in units of 
normalised background. The source of this systematic is as yet not fully understood. It may be linked to a 
`missing' telescope (CT2 is not present in large fraction of these data) which is uncovering inhomogeneities 
in trigger response correlating with certain telescopes, although this gradient does not align with the
CT-System geometry.  It is seen in these data that artificial removal of other telescopes for example, 
can also introduce further similar-scale effects which may or may not be related. Changes in average
skynoise across the FoV may also contribute. Care should 
therefore be taken when searching for sources in affected regions, employing alternative background 
estimates such as displacement backgrounds as a check. All background methods will of course also be 
affected by large-scale systematics of various types and so the above-mentioned problem is not particular to
the template model.
Extra correction terms could be added to the template model to accommodate such higher order
effects but this has not yet been explored in detail.
Experience has shown that for large exposures, higher order systematic effects can be averaged out
or considerably reduced by employing a variety of tracking positions, for example in a mosaic fashion. 
All of the example datasets presented here employed the so-called {\em wobble} mode in which 
only two tracking positions were  
alternated $\pm0.5^\circ$ in declination with respect to the primary source of interest. 
Investigation of higher order effects and their origin is presently continuing, in the context of a search
for TeV sources from the entire CT-System archive (P\"uhlhofer et al. \cite{Puehl:2}, 
Aharonian et al. \cite{Aharonian:4}). 

The estimation of energy spectra from source excesses requires that the background estimate be comprised 
of events with a similar distribution of estimated energies to that of events from the source region. 
The template background estimate however does not meet this criterion since the
\W values comprising the background estimate will differ in estimated energy from those in the
 $\gamma$-ray-like \W regime. Thus alternative background models should be used for spectral analysis.

\section{Conclusions}

Described in this paper is a new method to estimate the CR background in 2D searches for sources of 
TeV $\gamma$-ray emission with ground-based detectors. This so-called template background estimate, demonstrated 
here with HEGRA CT-System data, employs a subset of CR events normally rejected according to
the \C image shape parameter mean-scaled-width \W.
These template events spatially and temporally overlap with CR events considered `gamma-ray-like',
and thus no dedicated OFF source observations are required.
Applying the template model successfully to HEGRA CT-System data involves correction over of the field 
of view for: 
(1) differences in radial response and (2) differences in a zenith-correlated gradient, both
between the CR events of the two \W regimes.
A particularly useful feature of the template model is its ability to
compensate well for localised systematic changes in CR event density due to the presence of stars in the 
field of view. It is well suited to cases where many stars and TeV sources are present.
The template model applicability is limited by the presence of systematic
gradients in CR event density which are presently not corrected. All background models
will however suffer to various extent from weak, large scale systematics. Further
investigation of these aspects is ongoing.
Systematic uncertainties in the derived event excess of less than 4\% of the 
normalised background are achieved when searching for sources of size less than $\sim0.6^\circ$ radius in
many datasets.

It should be possible to extend the template model philosophy to analyses which use 
other means to reject CR background, for example
maximum likelihood methods, multi-dimensional cluster analyses, applied to either single or 
multi-telescope systems.  
Issues relating to systematic gradients over the FoV and star deficits
will be important for the next generation IACTs
such as H.E.S.S., VERITAS, CANGAROO III and MAGIC. In these new systems, CR event rates of order 500 Hz 
are expected, likely revealing higher-order systematic gradients in response after observation times 
significantly shorter than that required in HEGRA CT-System data.
Furthermore, the larger mirror areas of these future experiments will also increase considerably the number 
of stars that could cause detrimental systematics in the FoV response. The template   
background model could therefore be quite useful in 2D skymap construction, in future searches for 
TeV $\gamma$-ray sources.

\begin{acknowledgements} 
 HEGRA colleagues are thanked for advice with CT-System data analysis and valuable discussions
 on this topic (in particular Niels G\"otting, Dieter Horns and Gerd P\"uhlhofer). GPR
 acknowledges receipt of a von Humboldt Fellowship.
\end{acknowledgements}

\appendix

\section{Template Background Corrections}
  \label{sec:appendixa}
Summarised here are the two template background correction terms calculated specifically for the
HEGRA CT-System Tycho's SNR dataset (Fig.~\ref{fig:skymaps}).
\begin{enumerate}
 \item {\bf Radial Correction:} The radial event density
        for the entire dataset, centred on the various tracking positions may be fitted with 8th-order
        polynomials out to a radius $R=2.24^\circ$: (1) $p_s$ for the `gamma-ray-like' 
        (\W$<$1.1), and (2) $p_b$ for the template regime (1.3$<$\W$<$1.5):
        \begin{equation}
            p_{s,b} = \sum_{i=0}^{8} \wp_i \,\, (\theta^2_{\rm track})^i \nonumber
        \end{equation}
        where for Tycho's SNR data one fits:

        {\small
        \begin{center}
        \begin{tabular}{l|cc} \hline \hline
                &    $p_s$        &  $p_b$  \\ \hline   
        $\wp_0   $&$ 215.3\pm0.7$    & $815.3\pm0.9$ \\
        $\wp_1   $&$   17.0\pm0.8$     & $160.8\pm1.9$  \\
        $\wp_2   $&$ -148.7\pm0.1$     & $-606.2\pm0.8$ \\
        $\wp_3   $&$  148.00\pm0.02$   & $576.80\pm0.19$  \\
        $\wp_4   $&$  -62.340\pm$O($10^{-3}$) & $-243.60\pm0.04$  \\
        $\wp_5   $&$  7.987\pm$O($10^{-3}$) & $31.95\pm0.01$  \\
        $\wp_6   $&$  1.661\pm$O($10^{-4}$) & $6.117\pm$O($10^{-3}$)  \\
        $\wp_7   $&$  -0.5548\pm$O($10^{-5}$) & $-2.103\pm$O($10^{-4}$) \\
        $\wp_8   $&$  0.0418\pm$O($10^{-4}$) & $0.158\pm$O($10^{-3}$) \\ \hline \hline
        \end{tabular}
        \end{center}
        }

 \item {\bf Zenith-Correlated Correction:} A linear function is used to describe the ratio of
         event densities, as a function of the zenith angle difference $(\Delta z)$ 
         between the event and tracking positions, of the gamma-ray-like and template \W regimes,:
         \begin{equation}
           f(\Delta z) = (0.998\,\pm 0.006)\,\, +\,\, (0.898\,\pm 0.006)\, \Delta z \nonumber
         \end{equation} 
         A linear gradient of order 9\% per degree along the zenith axis is therefore present in the
         Tycho's SNR data. Separately, one can also fit functions to each \W regime; $f(\Delta z)_s$ and
         $f(\Delta z)_b$, and note that the function $f(\Delta z)=f(\Delta z)_s / f(\Delta z)_b$:
         \begin{eqnarray}
           f(\Delta z)_s = (0.999\,\pm 0.006)\,\, +\,\, (0.107\,\pm 0.006)\, \Delta z  \nonumber \\
           f(\Delta z)_b = (1.000\,\pm 0.006)\,\, +\,\, (0.015\,\pm 0.006)\, \Delta z  \nonumber
         \end{eqnarray}
        \noindent The zenith-correlated gradient therefore arises predominantly from CR events in the 
        gamma-ray-like \W regime.

        This gradient actually comprises two components: (1) the naturally expected event
        rate dependence upon zenith angle for a \C detection instrument, and (2) using in the construction
        of the \W parameter which employs a zenith-dependent scaling, the tracking zenith angle instead of
        that for each individual event. Here, the template model is demonstrated for a type of 
        worst-case-scenario, without removing component (2). Component (1) is usually of order few percent 
        per degree for data taken at zenith angles such in the Tycho's SNR dataset.
\end{enumerate}

It should also be noted that Eq. (~\ref{eq:signif}) takes no account of statistical errors in the 
normalisation $\alpha$, and the template correction terms. These errors are however generally less than one
percent (for obs. time $>10$ h) and can be 
neglected given the high significance levels to which these parameters are estimated.

\end{document}